\documentclass[aps,prb,twocolumn,showpacs,superscriptaddress,floatfix]{revtex4-1}
\usepackage{graphicx}
\usepackage{amssymb}

\usepackage{amsmath}
\usepackage{epsf}
\usepackage{psfrag,float}

\begin{document}

\title{Angular variation of the magnetoresistance of the superconducting ferromagnet UCoGe}

\author{T. V. Bay} \email{v.b.tran@uva.nl, tvanbay@gmail.com} \affiliation{Van der Waals - Zeeman Institute, University of Amsterdam, Science Park 904, 1098 XH Amsterdam, The Netherlands}
\author{A. M. Nikitin} \affiliation{Van der Waals - Zeeman Institute, University of Amsterdam, Science Park 904, 1098 XH Amsterdam, The Netherlands}
\author{T. Naka} \affiliation{National Institute for Materials Science, Sengen 1-2-1, Tsukuba, Ibaraki 305-0047, Japan}
\author{A. McCollam} \affiliation{High Field Magnet Laboratory, Radboud University, 6525 ED Nijmegen, The Netherlands}
\author{Y. K. Huang} \affiliation{Van der Waals - Zeeman Institute, University of Amsterdam, Science Park 904, 1098 XH Amsterdam, The Netherlands}
\author{A. de Visser} \email{a.devisser@uva.nl} \affiliation{Van der Waals - Zeeman Institute, University of Amsterdam, Science Park 904, 1098 XH Amsterdam, The Netherlands}

\date{\today}

\begin{abstract} We report a magnetoresistance study of the superconducting ferromagnet UCoGe. The data, taken on single-crystalline samples, show a pronounced structure at $B^* = 8.5$~T for a field applied along the ordered moment $m_0$. Angle dependent measurements reveal this field-induced phenomenon has an uniaxial anisotropy. Magnetoresistance measurements under pressure show a rapid increase of $B^*$ to 12.8~T at 1.0~GPa. We discuss $B^*$ in terms of a field induced polarization change. Upper critical field measurements corroborate the unusual S-shaped $B_{c2}(T)$-curve for a field along the $b$-axis of the orthorhombic unit cell.

\end{abstract}


\maketitle

\section{Introduction}

The intermetallic compound UCoGe belongs to the select group of superconducting ferromagnets~\cite{Huy-PRL-2007}. In this intriguing group of materials superconductivity develops in the ferromagnetic state at a temperature $T_s$ well below the Curie temperature $T_{C}$ for ferromagnetic ordering~\cite{deVisser-EMSAT-2010,Aoki-CRP-2011}. Moreover, below $T_s$, superconductivity and ferromagnetic order coexist on the microscopic scale. The superconducting ferromagnets discovered so far are UGe$_2$ (under pressure, Ref.\onlinecite{Saxena-Nature-2000}), URhGe (Ref.\onlinecite{Aoki-Nature-2001}), UIr (under pressure, Ref.\onlinecite{Akazawa-JPCM-2004}) and UCoGe (Ref.\onlinecite{Huy-PRL-2007}). The co-occurrence of ferromagnetism and superconductivity is at odds with the standard BCS (Bardeen-Cooper-Schrieffer) scenario for phonon-mediated spin-singlet superconductivity, since the ferromagnetic exchange field impedes spin-singlet Cooper pairing~\cite{Berk-PRL-1966}. Instead, alternative models have been proposed that exploit the itinerant nature of the ferromagnetic order where critical spin fluctuations, connected to a magnetic instability, mediate an unconventional, spin-triplet type of pairing~\cite{Fay-PRB-1980,Monthoux-Nature-2007}. Indeed, these uranium intermetallics, where the $5f$-electrons are delocalized, are all close to a magnetic instability that can be induced by mechanical pressure, chemical doping or an applied magnetic field~\cite{Pfleiderer-RMP-2009}. Unraveling the properties of superconducting ferromagnets might help to understand how spin fluctuations can stimulate superconductivity, which is a central theme for materials families as diverse as heavy-fermion, high-$T_s$ cuprate and iron-oxypnictide superconductors.

UCoGe crystallizes, just like URhGe, in the orthorhombic TiNiSi structure with space group $P_{nma}$ (Ref.~\onlinecite{Canepa-JALCOM-1996}). The coexistence of superconductivity and ferromagnetism was first reported by Huy \textit{et al.}~\cite{Huy-PRL-2007,Huy-PRL-2008}. High-quality single crystals with a typical residual resistance ratio, $RRR$ = $R$(300K)/$R$(1K), of 30 have a Curie temperature $T_{C} = 2.8$~K and show superconductivity with $T_s = 0.5 $~K. UCoGe is a uniaxial ferromagnet. The spontaneous magnetic moment, $m_0$, points along the $c$-axis and attains the small value of $0.07~\mu _B$ per U-atom in the limit $T \rightarrow 0$. Proof for the microscopic coexistence of superconductivity and ferromagnetic order is provided by $\mu$SR (muon spin relaxation and rotation)~\cite{deVisser-PRL-2009} and $^{59}$Co-NQR (nuclear quadrupole resonance)~\cite{Ohta-JPSJ-2010} experiments. Evidence for spin-triplet Cooper pairing has been extracted from the magnitude of the upper critical field $B_{c2}^{\perp}$ (measured with the external magnetic field directed perpendicular to $m_0$), which greatly exceeds the Pauli limit for spin-singlet superconductivity~\cite{Huy-PRL-2008,Aoki-JPSJ-2009}. The important role of spin-fluctuations in promoting superconductivity is established by the large anisotropy of the upper critical field, $B_{c2}^{\perp} \gg B_{c2}^{\parallel}$ (Refs.~\onlinecite{Huy-PRL-2008,Aoki-JPSJ-2009}). For $B \parallel m_0$ the magnetic transition becomes a cross-over, spin fluctuations are rapidly quenched and, accordingly, superconductivity is suppressed, while for $B \perp m_0$ spin fluctuations become more pronounced and superconductivity is enhanced. At the microscopic level, the close link between anisotropic critical magnetic fluctuations and superconductivity was recently put on a firm footing by $^{59}$Co-NMR (nuclear magnetic resonance)~\cite{Ihara-PRL-2010,Hattori-PRL-2012} and inelastic neutron scattering~\cite{Stock-PRL-2011}.

Yet another salient property of UCoGe is the unusual S-shaped curvature of the upper critical field for a field direction along the $b$-axis, $B_{c2}^{b}(T)$, which yields the large value of $\sim~18$~T when $T \rightarrow 0$ (Ref.~\onlinecite{Aoki-JPSJ-2009}). This field-reinforced superconductivity seems to be closely connected to a field-induced quantum critical point as a result of the progressive depression of the Curie temperature~\cite{Aoki-JPSJ-2009,Miyake-JPSJ-2008,Aoki-CRP-2011}. The peculiar response of the magnetic and superconducting phases to a magnetic field calls for a detailed investigation of the anisotropy in the magnetic, thermal and transport properties. Here we present an extensive angle dependent magnetotransport study on high-quality single crystals of UCoGe for fields directed in the $bc$- and $ac$-planes of the orthorhombic unit cell. We identify a pronounced maximum in the magnetoresistance for a sample with $RRR=30$ when the component of the field along the $c$-axis reaches a value $B^* = 8.5$~T. Measurements of $B^*$ as a function of pressure show a rapid increase of $B^*$ to 12.8~T at 1.0~GPa. The uniaxial nature of $B^*$ and its large pressure variation provide strong indications for a close connection to an unusual polarizability of the U and Co moments. Transport measurements around the superconducting transition in fixed magnetic fields $B \parallel b$ reveal our samples exhibit the S-shaped $B_{c2}$-curve  when properly oriented in the magnetic field.

\begin{figure}
\includegraphics[width=8cm]{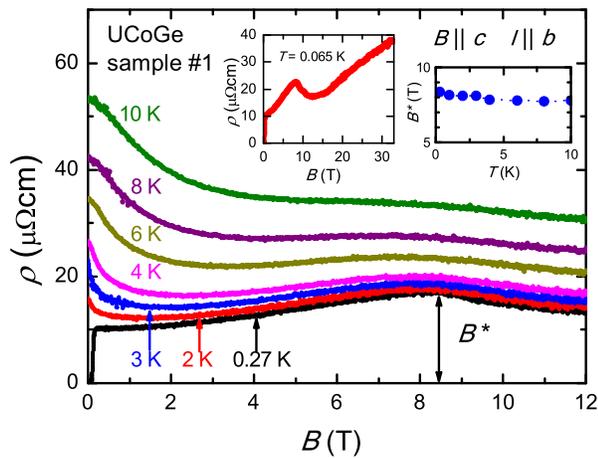}
\caption{(color online) Resistivity of UCoGe (sample \#1) as a function of the magnetic field $B \parallel c$ at temperatures of $0.27, 2, 3, 4, 6, 8$ and $10$~K, as indicated. The current was applied along the $b$-axis. Left inset: High-field magnetoresistance ($B \parallel c; I \parallel b$) up to $B= 33$~T at $T=0.065$~K. Right inset: $B^*$ as a function of temperature determined from the maximum in the magnetoresistance.}
\end{figure}

\section{Experimental}

Single cystals of UCoGe were prepared in a tri-arc furnace by the Czochralski technique as described in Ref.~\onlinecite{Huy-JMMM-2009}. Bar-shaped samples with typical dimensions $5 \times 1 \times 1$~mm$^3$ were cut from the crystals by means of spark erosion. Magnetotransport measurements were carried out on three samples with the current, $I$, along the $b$- and $c$-axis, with $RRR$-values of 30 (sample \#1) and 8 (sample \#2, sample \#3), respectively.
Experiments on sample \#1 were performed in a $^3$He refrigerator (Heliox, Oxford Instruments) in the temperature range 0.24-15~K and fields up to 14~T, and in a dilution refrigerator (Kelvinox M100, Oxford Instruments) in the temperature range 0.04-1.0~K and fields up to 16~T. In the Kelvinox the sample was mounted on a Swedish rotator for angle dependent measurements. In addition, experiments were carried out at the High Field Magnet Laboratory in Nijmegen in a dilution refrigerator and in fields up to 33~T. Sample \#2 was measured under pressure using a hybrid clamp cell attached to the cold plate of the  $^3$He refrigerator. Details of the pressure cell and calibration data are given in Ref.~\onlinecite{Slooten-PRL-2009}. Sample \#3 was measured in the $^3$He refrigerator at ambient pressure and $T=0.25$~K. The ac-resistivity data were acquired using a low-frequency ($f=16$~Hz) resistance bridge or a phase sensitive detection technique using a lock-in amplifier. Care was taken to prevent heating of the sample by selecting a low excitation current ($I=100~\mu$A).

\begin{figure}
\includegraphics[width=8cm]{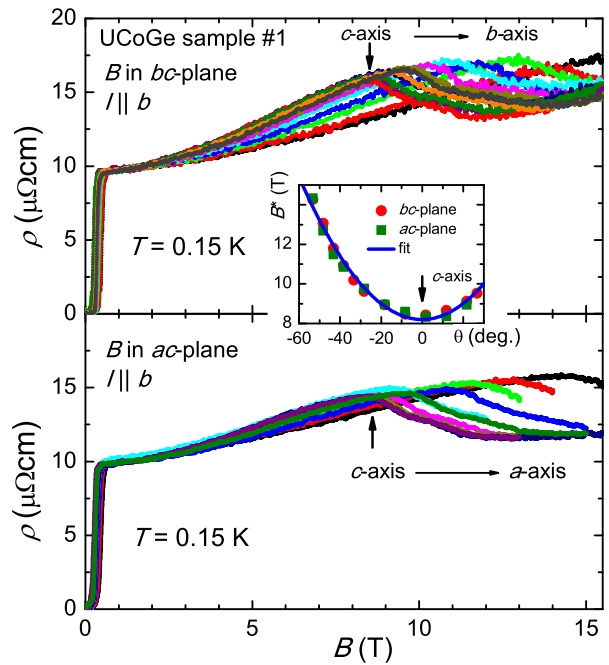}
\caption{(color online) Angular variation of the magnetoresistance of UCoGe (sample \#1) at $T=0.15$~K. Upper panel: field rotation in the $bc$-plane; $\theta$ = -58, -53, -48, -43, -38, -33, -28, -18, -8, 2, 12, 22 and 27 degrees, where $0^{\circ}$ corresponds to $B \parallel c$. Lower panel: field rotation in the $ac$-plane; $\theta$ = -53, -48, -43, -38, -28, -18, -8, 2, 12, 22 and 27 degrees. The current is always applied along the $b$-axis. Inset: $B^*$ as a function of $\theta$. The solid line represents $B^*(\theta) = B^*(0) / \cos \theta$. }
\end{figure}

\section{Results}

\subsection{Magnetoresistance}
In Fig.~1 we show the resistivity of UCoGe (sample \#1) as a function of the magnetic field applied along the $c$-axis. At the lowest temperature $T=0.27$~K ($< T_s)$ the initial steep rise signals the suppression of superconductivity at $B_{c2} = 0.2$~T. Next, $\rho (B)$ steadily increases and passes through a pronounced maximum at $B^* = 8.5$~T. Increasing the temperature shows that the maximum at $B^*$ is a robust property and can be identified up to at least 10 K. The temperature variation of $B^*$ is relatively weak as shown in the right inset of Fig.~1. We remark that the overall resistivity rapidly increases with temperature and in the normal phase the initial low-field magnetoresistance is negative. In the left inset we show data taken at $T=0.065$~K in strong magnetic fields up to 33~T. The maximum at $B^*$ is most pronounced. For fields exceeding 12~T the magnetoresistance displays a steady increase that leads to the large value of 40~$\mu\Omega$cm at the maximum field.

In order to investigate the magnetocrystalline anisotropy of $B^*$, we have measured the angle dependent magnetoresistance. The data taken in a dilution refrigerator at $T=0.15$~K are shown for a field rotation in the $bc$-plane and in the $ac$-plane in the upper and lower panels of Fig.~2, respectively. The major experimental observation is the steady upward shift of the maximum in $\rho(B)$ when the field is rotated away from the $c$-axis. The value $B^*(\theta)$ is proportional to $B^*(0)/\cos \theta$, where $\theta$ is the angle at which the field is tilted from the $c$-axis. This functional behavior is illustrated in the inset in Fig.~2 and holds for the $bc$- as well as for the $ac$-plane. For $\theta > 58^{\circ}$ the maximum in $\rho (B)$ falls outside the magnetic field range probed in the dilution refrigerator. We remark that the value of the maximum magnetoresistance $\rho ^{*}$ at $B^*$ is quasi field-angle independent. This tells us the angle dependent magnetoresistance data may be collapsed onto a single reduced curve $\rho / \rho ^{*}$ \textit{versus} $B/B^*$. We conclude the maximum in $\rho(B)$ takes place when the component of the magnetic field along the $c$-axis reaches $B^* = 8.5$~T. This confirms its uniaxial nature, just as for the ferromagnetic order. The suppression of superconductivity in the field-angle interval probed in Fig.~2 still takes place at a low value of $B_{c2}$~\cite{Aoki-JPSJ-2009}.

The pressure variation of $B^*$ was investigated for sample \#2 for $B \parallel I \parallel c$ in the pressure range 0.26-1.29 GPa in the $^3$He refrigerator. In this longitudinal configuration the field variation $\rho(B)$ is very different as shown in Fig.~3. This is confirmed by the longitudinal magnetoresistance of sample \#3, measured at $p=0$ (see right inset Fig.~3). After the initial steep rise, due to the suppression of superconductivity, $\rho(B)$ steadily decreases and shows a kink near 9.2~T rather than a maximum at ambient pressure. The field at which the kink appears identifies $B^*$. Under pressure $B^*$ increases rapidly ($\propto p^2$) up to 12.8~T at 1.0~GPa (see left inset Fig.~3 for $B^*$-values at $T= 0.25$~K and 1.0~K). The temperature variation is weak. At our highest pressure (1.29~GPa) $B^*$ falls outside the available field range in the $^3$He refrigerator.

\begin{figure}
\includegraphics[width=8cm]{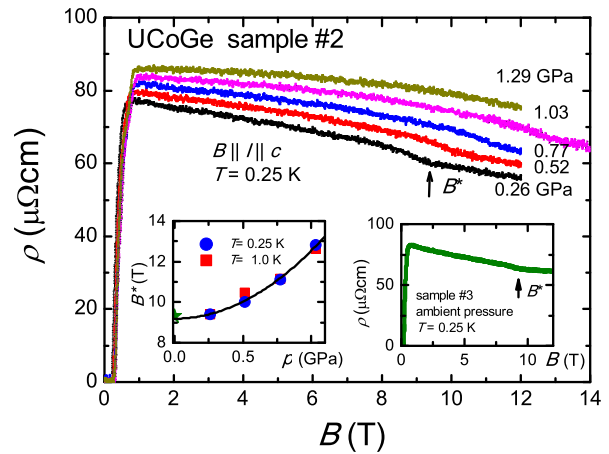}
\caption{(color online) Magnetoresistance of UCoGe (sample \#2) for $B \parallel I \parallel c$ at pressures of 0.26, 0.52, 0.77, 1.03 and 1.29 GPa as indicated. The temperature is $T=0.25$~K. Right inset: Magnetoresistance of sample \#3 for $B \parallel I \parallel c$ at ambient pressure at $T=0.25$~K. Left inset: $B^*$ as a function of pressure at $T=0.25$~K (circles) and $T=1.0$~K (squares). The value $B^*(0) =9.2~T$ at ambient pressure (symbol: star) is taken from sample \#3. The solid line is a fit to the data at $T=0.25$~K with $B^*(p) = B^*(0) + b p^2$ where $ b = 3.35$~T/GPa$^2$. }
\end{figure}

\subsection{Upper critical field}

\begin{figure}
\includegraphics[width=8cm]{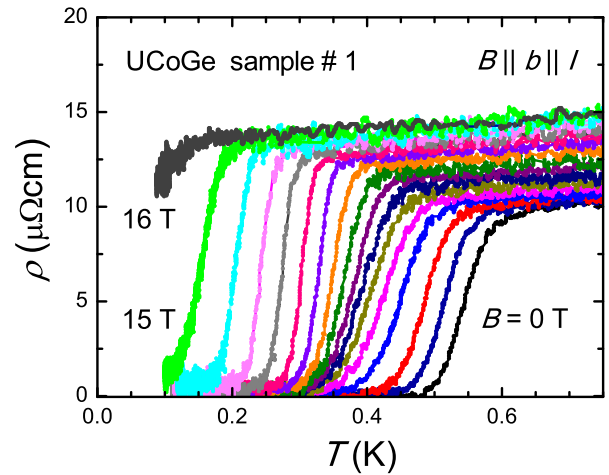}
\caption{(color online) Superconducting transition of UCoGe (sample \#1) measured by resistivity for $B \parallel b \parallel I$ in fixed magnetic fields from 0 to 16 T with steps of 1 T (from right to left).}
\end{figure}

The upper critical field $B_{c2}^{\perp}(T)$ for a field direction perpendicular ($B \parallel a$ or $B \parallel b$) to the ordered moment ($m_0 \parallel c$) is extremely sensitive to the precise orientation of the magnetic field~\cite{Aoki-JPSJ-2009}. In order to substantiate the unusual $B_{c2}^{b}$-behavior of our single crystals, we measured sample \#1 as a function of field orientation in the dilution refrigerator. Special care was taken to enable field rotation in the $bc$-plane. After fine tuning to $B \parallel b$ we measured the resistivity in fixed magnetic fields. The results are shown in Fig.~4. For $B=0$ the superconducting transition sets-in at 0.6~K and has a width $\Delta T_s = 0.1$~K. On applying a magnetic field, the superconducting transition progressively shifts to lower temperatures, and is still visible up to the highest field (16~T). Striking features are (\textit{i}) the different depression rate of $T_s$ in the field range 5-9~T, and (\textit{ii}) the narrowing of $\Delta T_s$ across the same field range. The upper critical field, determined by taking the midpoints of the transitions, is shown in Fig.~5. $B_{c2}^{b} (T)$ has an unusual curvature for $B > 4$~T and extrapolates to the large value of 17~T in the limit $T \rightarrow 0$, in good agreement with the results reported in Ref.~\onlinecite{Aoki-JPSJ-2009}. In the inset we show the large angular variation of $B_{c2}$ around $B \parallel b$ measured at $T=0.15$~K. For a tilt-angle of typically $2 ^{\circ}$ the upper critical field has diminished by a factor of 3 (Ref.~\onlinecite{Aoki-JPSJ-2009}).

\begin{figure}
\includegraphics[width=8cm]{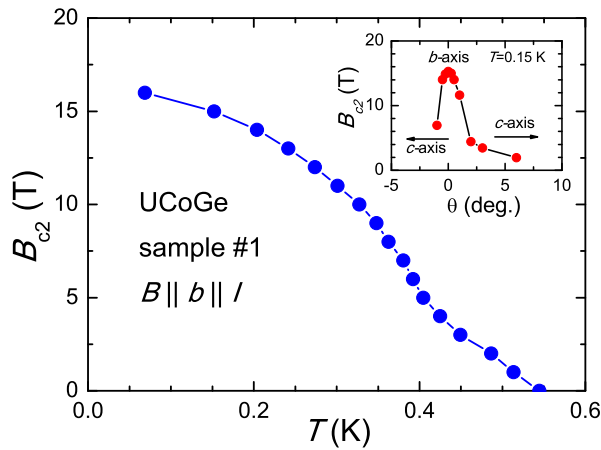}
\caption{(color online) Temperature variation of the upper critical field $B_{c2}(T)$ of UCoGe (sample \#1) measured for $B \parallel b$. Inset: Angular variation of $B_{c2}(T)$ in the $bc$-plane at $T = 0.15$~K.}
\end{figure}

\section{Discussion}

The major result from the angle dependent magnetoresistance measurements is the pronounced maximum at a field $B^*$, which occurs when the component of the magnetic field along the $c$-axis reaches a value of 8.5~T. The characteristic field $B^*$ is a robust property of our samples, but the shape of the magnetoresistance $\Delta \rho \equiv \rho(B) - \rho (0)$ is different for sample \#1 (with a maximum at $B^*$, Fig.~1) and samples \#2,\#3 (with a kink at $B^*$, Fig.~3). We remark that there are two obvious differences between the experiments. Firstly, the sample quality is very different as quantified by the residual resistivity value $\rho _0$ of 10 and $\sim 80~\mu \Omega$cm, respectively. Possibly, for samples \#2,\#3  magnetic disorder makes a large contribution to $\rho_0$, which can be reduced by the magnetic field resulting in a negative $\Delta \rho$. The second difference is the measurement geometry, \textit{i.e.} transversal ($B \parallel c,~I \parallel b$ for sample \#1) \textit{versus} longitudinal ($B \parallel c \parallel I$ for samples \#2,\#3) magnetoresistance, since the Lorentz force on the current in general leads to more scattering and a positive $\Delta \rho$. Moreover, in the transverse configuration the charge carriers will scatter more effectively from the magnetic moments and the predominantly longitudinal spin fluctuations, compared to the longitudinal configuration where the current and $m_0$ are aligned. These differences in sample quality and measurement geometry might also explain the slightly different $B^*$-values extracted from Fig.~1 ($8.5$~T) and Fig.~3 ($9.2$~T) at ambient pressure. The large variation of $\Delta \rho$ with the $RRR$-value and geometry is uncommon and its understanding is highly relevant in view of the strongly anisotropic magnetic properties of UCoGe.

Measurements in the transverse geometry with $B \parallel c$ have not appeared in the literature so far, while longitudinal ($c$-axis) magnetoresistance data have been reported on two samples of different quality: (\textit{i}) a magnetoresistance trace taken on a sample with $RRR=30$ at $T=0.04$~K shows a weak initially positive $\Delta \rho$ with a small structure near $B^* \approx 9$~T and three additional kink-like  features in the field range 17-30~T (Ref.~\onlinecite{Aoki-JPSJ-2011}), and (\textit{ii}) the magnetoresistance of a sample with $RRR=5$ has an overall negative $\Delta \rho$ with a kink at $B_k$ or $B^* \approx 9$~T (Ref.~\onlinecite{Steven-APL-2011}), as in our Fig.~3. In the latter study the angular variation of $B^*$, measured at $T=0.04$~K by tilting the field from the $c$-axis towards an arbitrary direction in the $ab$-plane, was also found to follow the $B^*(\theta =0)/\cos \theta$-law.

An appealing scenario that has been put forward to explain the change in magnetoresistance at $B^*$ is a ferro-to-ferrimagnetic transition~\cite{Steven-APL-2011}. This proposal is largely based on a recent polarized neutron diffraction experiment on UCoGe carried out for $B \parallel c$~\cite{Prokes-PRB-2010}. In low magnetic field (3~T) the small ordered moment $m_0$ is predominantly located at the U atom, but in a large field of 12~T a substantial moment, antiparallel to the U moment, is induced on the Co site. This unusual polarizability of the Co $3d$ orbitals may give rise to a field-induced ferri-magnetic-like spin arrangement. Support for this scenario was obtained by field-dependent ac-susceptibility data~\cite{Steven-APL-2011} which exhibit a maximum near $B^*$. Recently, the dc-magnetization $M(B)$ was measured at $T=1.5$~K in pulsed magnetic fields up to 52~T~\cite{Knafo-PRB-2012}. For $B \parallel c$ the data do not show a clear sign of a (meta)magnetic transition, however, a weak structure appears near $B^*$ in the derivative $dM/dB$, and a second change of slope occurs near 23.5~T. Sensitive torque cantilever experiments might be helpful to resolve the possibly anomalous behavior of the magnetization around $B^*$. Further arguments in favor of a magnetic transition are: (\textit{i}) the uniaxial (Ising-type) behavior of the ferromagnetic order is reflected in $B^*$, and (\textit{ii}) the pressure variation of $B^*$ (see Fig.~3) is large and has a magnitude comparable to the pressure dependence of $T_{C}$~\cite{Slooten-PRL-2009} assuming 1~K $\approxeq 1.5$~T per $\mu_B$ (the critical pressure for the suppression of ferromagnetic order is 1.4~GPa). In this scenario the pressure increase of $B^*$ may be related to the reduced polarizability of the Co moment under pressure.

Another possible origin of the structure in $\Delta \rho$ near $B^*$ is a Lifshitz transition, \textit{i.e.} a field-induced topological change of the Fermi surface. Notably it has been suggested that the  multitude of small kink-like features observed in $\Delta \rho$ for $B \parallel c \parallel I$ at $T=0.04$~K could hint at a Fermi surface reconstruction~\cite{Aoki-JPSJ-2011}. Quantum oscillations have been reported for UCoGe for $B \parallel b$ but could not be detected for a field direction along or close to the $c$-axis. A second indication for the possibility of a field-induced Fermi surface modification comes from thermoelectric power data~\cite{Malone-PRB-2012}, which show two pronounced peaks at 11.1 and 14.6~T for $B \parallel b$. While the former peak is associated with field-reinforced superconductivity, the latter peak and the ensuing sign change of the thermopower provide evidence for a topological change of the Fermi surface. In the related material URhGe the field-induced disappearance of a small Fermi-surface pocket was recently demonstrated by quantum oscillations measurements~\cite{Yelland-NatPhys-2011}.

Overall, the Ising-like nature of the ferromagnetic ground state results in a complex magnetotransport behavior. Moreover, the magnetization~\cite{Huy-PRL-2008,Hardy-PRB-2011,Knafo-PRB-2012}, thermal expansion~\cite{Gasparini-PRB-2010}, thermoelectric power~\cite{Malone-PRB-2012} and thermal conductivity~\cite{Taupin-PRB-2014} all have a strong magnetocrystalline anisotropy, which makes it difficult to unravel the behavior of UCoGe. However, at the positive side, it is the strong anisotropy that results in longitudinal ferromagnetic fluctuations that play a major role in inducing spin-triplet superconductivity~\cite{Hattori-PRL-2012}. A greater understanding of the anisotropy is therefore likely to be important for our understanding of the superconductivity.

The unusual superconducting behavior is demonstrated by the $B_{c2}(T)$ curve reported in Fig.~5. We recall the upward curvature for $B > 4$~T and the large value of 17~T in the limit $T \rightarrow 0$. The sample (\#1) used here comes from the same single-crystalline batch as used in our first measurements of the upper critical field (Ref.~\onlinecite{Huy-PRL-2008}), where $B_{c2}(0)$ was found to reach a value of 5~T for $B \parallel b$. This discrepancy can now be attributed to a small misorientation of $\sim 2^{\circ}$ (see the inset in Fig.~5). The precise orientation of the sample with respect to the magnetic field direction remains an absolutely crucial feature for the behavior of this material. The field-reinforced superconductivity appears to be connected to critical spin fluctuations associated with a field-induced quantum critical point, where the latter is reached by the suppression of the Curie temperature in strong magnetic fields for $B \perp m_0$~\cite{Mineev-PRB-2011}. A second, more recently, proposed cause for the field-reinforced superconductivity is a Lifshitz transition~\cite{Yelland-NatPhys-2011,Taupin-PRB-2014}. Finally, we mention the progress made in modeling the intricate and anisotropic $B_{c2}(T)$ of UCoGe by a strong-coupling Eliasbergh model exploiting the Ising-type spin fluctuations~\cite{Tada-JPSJ-2011} and by the completely broken symmetry scenario for parallel-spin $p$-wave superconductors~\cite{Lorscher-PRB-2013}.

\section{Summary}

In summary, we have presented an extensive angle dependent magnetoresistance study on single crystals of UCoGe for fields directed in the $bc$- and $ac$-planes of the orthorhombic unit cell. We pinpoint a pronounced structure in the magnetoresistance, which takes place when the component of the field along the $c$-axis reaches a value $B^* = 8.5$~T. This behavior is very pronounced for transverse measurement geometry and rather weak for longitudinal geometry. Measurements of $B^*$ as a function of pressure show a rapid increase of $B^*$ to 12.8~T at $p= 1.0$~GPa. The uniaxial nature of $B^*$ and its large pressure variation are consistent with the interpretation that the change in the magnetoresistance regime at $B^*$ is related to an unusual polarizability of the U and Co moments. Transport measurements in fixed magnetic fields confirm the unusual S-shaped $B_{c2}(T)$-behavior after carefully aligning the sample along the field $B \parallel b$. In order to further unravel the intriguing properties of UCoGe, notably with respect to the close connection between field-induced phenomena, such as a quantum critical point or Lifshitz transition, and superconductivity, it requires an unremitted research effort to probe the strongly anisotropic thermal, magnetic and transport properties with help of high-quality single crystals.

Acknowledgements $-$ T.V. Bay acknowledges support of the Vietnamese Ministry of Education and Training. This work was part of the research program of FOM (Dutch Foundation for Fundamental Research of Matter). Part of this work was carried out at the HFML-RU/FOM, member of the European Magnetic Field Laboratory (EMFL).

\end{document}